TITLE:
Detecting interspecific and geographic differentiation patterns in two interfertile oak species (*Quercus petraea* (Matt.) Liebl. and *Q. robur* L.) using small sets of microsatellite markers

AUTHORS:
C. Neophytou[1, 2], F.A. Aravanopoulos[3], S. Fink[2] and A. Dounavi[1]

POSTAL ADDRESSES:
Forest Research Institute (FVA) Baden Württemberg,
Wonnhaldestr. 4
79100 Freiburg
Germany

Chair of Forest Botany
Faculty of Forest and Environmental Sciences
Albert-Ludwigs University of Freiburg
Bertoldstr. 17
79085 Freiburg
Germany

Laboratory of Forest Genetics and Tree Breeding
Faculty of Forestry and Natural Environment
Aristotle University of Thessaloniki
Po. Box. 238 Thessaloniki
Greece

AUTHOR FOR CORRESPONDENCE:
Charalambos Neophytou

Address:
Forest Research Institute (FVA) Baden Württemberg,
Wonnhaldestr. 4,
79100 Freiburg
Germany

Tel. number: +49 761 40 18 159
Fax number: +49 761 40 18 333
E-mail address: charalambos.neophytou@forst.bwl.de

ABSTRACT

Genetic analysis was carried out in order to provide insights into differentiation among populations of two interfertile oak species, *Quercus petraea* and *Quercus robur*. Gene flow between the two species, local adaptations and speciation processes in general, may leave differential molecular signatures across the genome. Three interspecific pairs of natural populations from three ecologically different regions, one in central Europe (SW Germany) and two in the Balkan Peninsula (Greece and Bulgaria) were sampled. Grouping of highly informative SSR loci was made according to the component of variation they express – interspecific or provenance specific. 'Species' and 'provenance discriminant' loci were characterized based on $F_{ST}$s. Locus specific $F_{ST}$s were tested for deviation from the neutral expectation both within and between species. Data were then treated separately in a Bayesian analysis of genetic structure. By using three 'species discriminant' loci, high membership probability to inferred species groups was achieved. On the other hand, analysis of genetic structure based on five 'provenance discriminant' loci was correlated with geographic region and revealed shared genetic variation between neighbouring *Q. petraea* and *Q. robur*. Small sets of highly variable nuclear SSRs were sufficient to discriminate, either between species or between provenances. Thus, an effective tool is provided for molecular identification of both species and provenances. Furthermore, data suggest that a combination of gene flow and natural selection forms these diversity patterns. 'Species discriminant' loci might represent genome regions affected by directional selection, which maintains species identity. 'Provenance specific' loci might represent genome regions with high interspecific gene flow and common adaptive patterns to local environmental factors.

1. INTRODUCTION

Since Darwinian times, oaks have served as a model genus for studying evolutionary processes and speciation. High adaptability and high levels of interspecific gene flow were among the features which significantly contributed to the genesis of several hundreds of species, subspecies and ecotypes (Kleinschmit, 1993; Petit et al., 2004; Stebbins, 1950). Species identification and characterization have often been a challenge since interspecific barriers are very weak. In recent years, a growing number of studies support the notion that differentiation is restricted to some limited genomic hot-spots, while the rest of the genetic information remains shared between related taxa (Bodénès et al., 1997b; Curtu et al., 2007a; Muir and Schlötterer, 2005).

*Q. petraea* and *Q. robur* are the most important oak species in central Europe, both economically and ecologically. They are interfertile and have widely overlapping distribution range. Within this range, *Q. robur* grows on humid soils with high nutrient availability, while *Q. petraea* rather occupies dry and more acidic sites (Rushton, 1979). However, the two species often occur in sympatry and hybridization under natural conditions is common (Bacilieri et al., 1996; Curtu et al. 2007b; Gugerli et al., 2007; Lepais et al., 2009; Rushton, 1979, Streiff et al., 1999). Extensive studies using cpDNA markers point to chlorotypes being shared between *Q. petraea*, *Q. robur* and other related species. This supports historical introgressive hybridization, which has led to frequent cytoplasmic exchanges through pollen swamping (Petit et al., 1997; Petit et al., 2002). Correlation between cpDNA haplotypes and variation of nuclear DNA marker loci could not be found (Finkeldey and Matyás, 2003; Kremer et al., 2002). This suggests that, after some generations of backcrosses, cpDNA of one species, which had originally colonized a site, could be captured by the other. In this case, hybridization has been described as a mechanism of invasion (Petit et al., 2004). At the nuclear level, comparative studies between *Q. petraea* and *Q. robur* have not revealed diagnostic alleles. Differentiation between the two species is restricted to allele frequencies, which differ significantly for some loci, as demonstrated by a large number of studies (e.g. Gömöry et al., 2001; Muir et al., 2000; Muir and Schlötterer, 2005; Scotti-Saintagne et al., 2004a; Zanetto et al., 1994). In contrast, differentiation within species is generally lower, but also varies from locus to locus and exceeds the interspecific one in certain large-scale studies (Bodénès et al., 1997a; Mariette et al., 2002).

With the development of numerous molecular markers during the last two decades, it has been possible to describe differentiation within and between *Q. petraea* and *Q. robur* with increasingly higher resolution. A common finding of many studies is that only a small fraction among hundreds of marker loci is highly differentiated, while for the rest of them differentiation is low or even non-significant (e.g. Bodénès et al., 1997b; Scotti-Saintagne et al., 2004a). High differentiation at $F_{ST}$ outlier loci (i.e. with interspecific $F_{ST}$ values which are significantly higher than those expected under selectively neutral conditions) arises from the fact that one allele occurs with exceptionally high frequency tending to fixation in one species, while the same locus remains highly polymorphic in the other (Muir et al., 2000; Muir and Schlötterer, 2005). This might be a result of selection acting upon genes, which are closely linked to the marker loci. Indeed, association studies reveal linkage of $F_{ST}$ outlier loci with QTLs that differentiate *Q. petraea* and *Q. robur*. For example, the nuclear microsatellite QrZAG96 is linked with a QTL coding for petiole length as a ratio of the total leaf length, which is a morphological trait discriminating *Q. petraea* from *Q. robur* (Saintagne et al., 2004). Additionally, recent studies indicate that the same differentiation pattern occurs between other pairs or groups of taxonomically related oak species. For instance, Curtu et al. (2007a) describe similar highly differentiated SSR loci distinguishing among *Q. petraea*, *Q. robur*, *Q. pubescens* and *Q. frainetto*. There is increasing evidence that directional selection at very limited genome regions following hybridization accounts for maintaining species integrity, while species are able to exchange genetic variants in other genome regions (Lexer et al., 2006, Petit et al., 2004). The same phenomenon has been reported for several interfertile plant and some animal species as well (Barton and Gale, 1993; Via and West, 2008).

In a study of *Q. petraea* Kremer et al. (2002) compared phenotypic and genetic divergence (RAPD markers) with cpDNA lineages among different provenances. They suggest that after recolonization of Central Europe, pollen mediated gene flow had a homogenizing effect between the separate gene pools of glacial refugia. This is reflected in the variation of RAPDs and the lack of correlation with cpDNA lineages. In contrast, subsequent selection for several phenotypic traits led to local adaptations and geographic variation trends. For instance, phenological traits such as bud burst display strong altitudinal and latitudinal variation clines corresponding to ecological gradients (Ducousso et al., 1996). This process of local adaptation apparently affects specific genomic regions. A relatively limited amount of research has dealt with a comparison of interspecific and intraspecific molecular differentiation in *Q. petraea* and *Q. robur*. A common finding is that, as expected, interspecific variation exceeds the intraspecific one (Bodénès et al., 1997a; Mariette et al., 2002; Zanetto et al., 1994). However, patterns of intraspecific variation differ among loci. In their study with SCARs, Bodénès et al. (1997a) compared differentiation within and between the two species using eight interspecific pairs of neighboring stands located across Europe. Interestingly, they described one locus (B11-1500) exhibiting extremely low interspecific differentiation between neighbouring populations, while differentiation among populations from different geographic locations was significant. A combination of interspecific gene flow and adaptation to heterogeneous site conditions may account for this result.

In our study, we aimed to capture the genetic imprints of the aforementioned evolutionary processes. By choosing neighbouring stands of *Q. petraea* and *Q. robur* from three different regions along an ecological gradient we aimed to observe the differential effects of gene flow and local adaptations on several genomic loci. We aimed to partition the loci according to the component of genetic variation they include. First, loci with consistently high interspecific differentiation may represent genomic regions which reflect species' ecological requirements and the respective adaptive profiles. Second, loci which vary strongly among regions, but show low interspecific differentiation at the local scale may indicate adaptations to the regional conditions, which are common between the two species and transferable through gene flow. Third, at selectively neutral loci gene flow is expected to play a homogenizing role resulting to loose genetic structures among regions or between species. The null hypothesis of no genetic structure among populations was tested by using different sets of loci.

2. MATERIALS AND METHODS

2.1. Sample collections

For this purpose, we sampled phenotypically pure autochthonous populations of *Q. petraea* and *Q. robur* from Central Europe and the Balkan Peninsula, along an ecological gradient with increasing aridity towards south. Old growth even aged stands were sampled in all cases. Trees were chosen to be at least 30 m apart from each other in order to avoid capturing family structures. A total of 48 samples were collected from each stand. Depending on the season, leaf material or buds were sampled from each individual.

The specific sampling locations were the following: For *Q. petraea* sampling locations included Emmendingen (48°08′N, 7°51′E) in Germany, Kavakliika (42°29′N, 25°13′E) in Bulgaria and Mount Holomon (40°28′N, 23°32′E) in northern Greece. For *Q. robur* the respective locations were Emmendingen (48°17′N, 7°42′E), Chirpanskata Gora (42°13′N, 25°16′E) and Lake Doirani (41°15′N, 22°46′E). Sampling locations are shown in Figure 1. In order to estimate aridity we used both Thornthwaite's precipitation effectiveness index (Thornthwaite, 1931) and De Martonne's aridity index (De Martonne, 1926), which are based on precipitation and mean annual temperatures. A thorough discussion about the use of climatic parameters and indices in plant geography is included in Tuhkanen (1980). The respective formulas were used to calculate values for each stand: (1) PE= $\Sigma_{n=1}^{12}[115*[p_n/(t_n-10)]^{10/9}]$, where PE= Thornthwaite's precipitation effectiveness index, $p_n$= mean monthly precipitation in inches, T= mean monthly temperature in °F, (2) H= P/(T+10), where H= de Martonne's aridity index, P= mean annual precipitation in mm, T= mean annual temperature in °C. The latter index can be used both for annual or seasonal data (Tuhkanen, 1980). We made the respective calculations by using means of the months July-September in order to estimate summer drought. In all cases, lower values indicate increased aridity. For the calculations we used climatic data from meteorological stations in the vicinity of each stand for the period 1961-1990. In general, data show an increasing aridity towards south, especially concerning the summer months (Table 1).

*2.2. Laboratory procedures*

Leaf or bud tissue was first freeze-dried and then DNA was extracted using the DNeasy 96 extraction kit (Qiagen). Subsequently, PCR was carried out using primers for twenty microsatellite loci. Marker loci included MSQ4 and MSQ13, initially described in *Q. macrocarpa* (Dow et al., 1995); QM50-3M in *Q. myrsinifolia* (Isagi and Suhandono, 1997); ssrQpZAG 1/5, 9, 15, 16, 46, 104 and 110 in *Q. petraea* (Steinkellner et al., 1997) and ssrQrZAG 7, 11, 30, 39, 87, 73, 75, 96, 101 and 112 in *Q. robur* (Kampfer et al., 1998). PCR protocols were based on Kampfer et al. (1998) with some modifications. PCR programs included an initial denaturation step at 95°C lasting 8 min, 23 cycles of 94°C for 15 s, an annealing step at temperature depending on primer for 15 s, an elongation step at 72°C for 15 s and ten additional cycles with reduced denaturation temperature (89°C). No final elongation was performed. Allele scoring was carried out by means of capillary electrophoresis using an ABI PRISM 3100 genetic analyzer (Applied Biosystems). Software GeneMapper v4.0 (Applied Biosystems) was used for genotyping. Allele bins were set by the software and were then proofed for consistency and corrected manually if necessary. In order to check for systematic genotyping errors we repeated fragment analysis by applying four samples from each one of the six study populations at the same plate.

*2.3. Data Analysis*

First, we controlled our data for possible scoring errors and occurrence of null (non-amplifiable) alleles using the software Microchecker (Van Oosterhout et al., 2004). Scoring errors tested, included large allele drop-off and error due to stuttering. In the first case, large-sized alleles are poorly amplified during PCR, thus leading to homozygote excess of small-sized alleles. Scoring errors due to stuttering result in a deficiency of heterozygote individuals differing by a single repeat. Finally, in the case of 'null alleles', there is an overall significant homozygote excess. One thousand randomizations of alleles within each locus and population were applied for the tests. Frequencies of such 'null alleles' were calculated according to Van Oosterhout et al. (2006). When tests supported presence of null alleles in more than two populations, loci were removed from further analyses, since they could lead to an overestimation of the homozygosity in a population.

Subsequently, we calculated population diversity measures. Observed and expected heterozygosity were calculated using the GenAlEx 6.1 software (Peakall and Smouse, 2006). We additionally used the FSTAT software (Goudet, 1995) to calculate allelic richness after rarefaction for each population and locus, which is a diversity measure independent of the population size (Petit et al., 1998). Rarefaction size was 31 corresponding to the minimum number of successful amplifications at one locus within a population (in our case QrZAG39 in the Greek *Q. petraea* population). Due to low amplification success, data from locus QpZAG46 in the Bulgarian populations of both *Q. petraea* and *Q. robur* were not included into allelic richness calculation. By using the same software we computed inbreeding coefficients ($F_{IS}$) for each locus and population and tested significance by performing 10000 randomizations of alleles among individuals within samples.

In order to reveal differentiation patterns, we calculated population pairwise $F_{ST}$s (Weir and Cockerham, 1984) by using the Arlequin 3.1 software (Excoffier et al., 2005). We measured interspecific differentiation by pooling all *Q. petraea* populations on one hand and all *Q. robur* populations on the other and by calculating a pairwise $F_{ST}$. We additionally measured differentiation among provenances within each species, by computing $F_{ST}$s among populations within each species. We tested significance by using 10000 permutations of individuals between the compared populations. Additionally, we compared the differentiation at our analyzed loci to its neutral evolutionary expectation. In the lack of natural selection, the distribution of $F_{ST}$ values is strongly related to the expected heterozygosity of a locus (Beaumont and Nichols, 1996). Thus, we applied the software Fdist2 to define a null distribution including 95% of the expected $F_{ST}$ values as a function of heterozygosity. Based on the average $F_{ST}$ of our analyzed loci, the software generated 10000 iterations of $F_{ST}$ and heterozygosity values. We then compared the observed $F_{ST}$ and $H_e$ of our loci to the modeled values and tested their significance. Three separate analyses were carried out: (1) in order to evaluate intraspecific differentiation we analyzed the data among populations separately within each species; (2) in order to evaluate interspecific differentiation we treated each species as one population consisting of all three sampled stands.

Linkage disequilibrium tests were performed in each population using the Arlequin 3.1 software. For testing linkage disequilibrium, a likelihood-ratio test with implementation of the EM algorithm was carried out. A total of 1000 permutations of alleles between individuals at each single locus were performed.

Population genetic structure was studied by using software Structure 2.2 (Pritchard et al., 2000). Based on the previously calculated $F_{ST}$s we examined: (a) Population structures derived from all loci used, (b) Population structures derived from loci where interspecific $F_{ST}$ values were larger than both within-species $F_{ST}$ values ('species discriminant loci'), (c) Population structures derived from loci with both within-species $F_{ST}$ values larger than $F_{ST}$ values between species ('provenance discriminant loci').

The Structure software 2.2 uses a Markov chain – Monte Carlo procedure to infer unstructured subpopulations, without considering prior classification within the sample. Proportions of membership of each individual to each one of K modeled subpopulations are calculated based on its genotype. For our analysis we chose the admixture model which assumes that individuals may have mixed ancestry and could thus be assigned to more than one subpopulation, given that *Q. petraea* and *Q. robur* are interfertile (Falush et al. 2003). Furthermore, we used the correlated allele frequency model. For the main data analysis we applied 100,000 updates of the Markov chain and 100,000 MCMC iterations. We ran our data for K=1…6. Ten runs were performed for each K. Maximum posterior probability (lnP(D)) and minimum deviation between runs were used to infer ΔK, a statistic based on the rate of change of lnP(D) between successive K values and its variation among runs. ΔK corresponds to the second order rate of change of lnP(D) for a given value of K and more accurately detects the uppermost hierarchical level of structure, providing an accurate estimation of population clustering (Evanno et al., 2005).

3. RESULTS

*3.1. Population diversity and differentiation*

High allelic richness and heterozygosity were generally observed in our samples. Overall expected heterozygosity varied between 0.781 and 0.815 in the populations of *Q. petraea* and between 0.814 and 0.817 in the populations of *Q. robur*. Allelic richness per locus ranged between 2.00 and 28.15 among populations of *Q. petraea* and between 4.29 and 28.08 among populations of *Q. robur* (Tables 2, 3). Significant inbreeding coefficients ($F_{IS}$) due to homozygote excess were observed in several cases. This was mostly due to 'null alleles', as supported by the results from the Microchecker software. There was no evidence about large allele drop-out or scoring errors due to stuttering. Loci MSQ4, QM50-3M, QpZAG46, QrZAG39, 73 and 75 were rejected from further analyses, since test for 'null alleles' was significant in more than two populations. 'Null allele' frequency at these loci varied between 5 and 33% (Tables 2, 3). Additionally, low amplification success resulted in a high number of missing values in the Bulgarian populations of both *Q. petraea* and *Q. robur*. In spite of many optimization trials including change of PCR conditions and re-extraction of the samples, we could genotype only a limited number of individuals from these populations. The same locus presented limited amplification in other populations, too (e.g. only 69% of successful amplifications in the Greek *Q. robur*). Similar problems have been observed by other laboratories as well (S. Nowak, personal communication).

The 14 remaining loci presented generally non-significant $F_{IS}$ values with a few local departures. Three loci exhibited high frequency of one allele and very low diversity in one species, whereas they remained highly variable in the other (Figure 1). For instance, locus QrZAG112 showed generally low diversity in *Q. petraea* with expected heterozygosity values varying between 0.173 and 0.366 and allelic richness between 2.00 and 3.65 (Table 2). The respective values in *Q. robur* were 0.811-0.836 (expected heterozygosity) and 13.39-17.34 (allelic richness; Table 3). A similar case can be observed at locus QrZAG30. On the other hand, *Q. robur* displays low diversity at locus QrZAG96 ($H_e$= 0.266-0.391 and A= 4.29-7.10), while *Q. petraea* is more diverse ($H_e$= 0.766-0.870 and A= 13.92-16.98).

$F_{ST}$ values between the two species reflect these differences in diversity. Specifically, loci QrZAG30, QrZAG96 and QrZAG112 exhibit high levels of interspecific differentiation (Table 4). $F_{ST}$ values between species are several times larger than among populations of the same species for these loci. Therefore, henceforth we call these loci 'species discriminant'. Allele frequency diagrams indicate that high levels of interspecific differentiation are due to a high degree fixation of a specific allele in one species, while diverse frequency patterns are kept in the other (Figure 2).

Conversely, for another set of five loci we observed higher $F_{ST}$ values among populations of the same species, than between species in general (Table 4). For example, locus QrZAG101 displays a non-significant value of 0.002 when the two species are compared. When populations of each species are treated separately $F_{ST}$s increase to 0.058 in *Q. petraea* and 0.070 in *Q. robur*. An examination of allele frequencies at this locus reveals that allele "153" occurs in high frequencies in both species in the two German populations, while its frequency decreases towards Greece irrespective of the species (Figure 3). Moreover, frequency gradients for this and for other alleles can be observed along the three areas and patterns are mostly shared between the two species for a specific sampling region. Loci with similar behaviour
were QpZAG1/5, 15, 110 and QrZAG87. Hereafter, we call all five loci with higher intraspecific than interspecific differentiation 'provenance discriminant'.

Subsequently, we tested the aforementioned $F_{ST}$-values against the evolutionary neutral expectation within and between species. The observed $F_{ST}$ value exceeded the neutral expectation significantly in only one case. In particular, locus QrZAG96 possessed a probability of 0.999 of being equal or larger than the expected value when we compared the two species (Table 5). Two more loci, QrZAG30 and QrZAG112 displayed high P-values between species (0.952 and 0.935 respectively) although not being outliers. No $F_{ST}$-outliers with P-values higher than 0.975 were found at the intraspecific level (by testing $F_{ST}$-values among populations within species). In two cases - QpZAG9 in *Q. petraea* and QrZAG104 in *Q. robur* - differentiation was lower than the expected under selectively neutral conditions with P-values being less than 0.025. Low differentiated loci were more common at the interspecific level. P-values lower than 0.025 were found at several loci (Table 5). All five 'provenance discriminant' loci were less differentiated than expected under selective neutral conditions. At the same time, these loci were more differentiated within species with P-values exceeding 0.8 in the case of QrZAG87 (in both species), QrZAG101 (in *Q. robur*) and QrZAG15 (in *Q. petraea*).

Finally, the test for linkage disequilibrium supports a lack of any linked loci pair across all populations. A limited number of linkage disequilibria were locally observed (results not shown). Hence, all fourteen loci were used for the Bayesian analysis of genetic structures.

*3.2. Analysis of genetic structure in Q. petraea and Q. robur*

Initially, all loci were jointly used for calculating proportions of membership to assumed K subpopulations. The rate of change of data posterior probability (lnP(D)) between successive runs was maximum for K= 2, giving the highest posterior probability of K (Figure 4). Two homogenous subpopulations, corresponding to the two species, were detected, when we set K= 2. Proportions of membership of each one of the original populations to the modeled subpopulations were high and varied between 0.977 and 0.988 for the *Q. petraea* populations and between 0.970 and 0.990 for *Q. robur* (Figure 5). In the whole sample (288 individuals) five trees presented individual membership proportions to an inferred cluster lower than 0.85 (two *Q. petraea* and two *Q. robur* from Bulgaria, and one *Q. robur* from Germany). Three designated *Q. robur* and seven designated *Q. petraea* individuals showed slightly admixed genotypes (membership proportion to the respective cluster between 0.85 and 0.95; results not shown). The respective number for *Q. petraea* was seven. By setting K= 3, clustering pattern was not consistent among the ten runs performed. For K= 4, within species structure was observed for both *Q. petraea* and *Q. robur*. The two German populations were genetically distinct (proportion of membership to separate groups 0.962 and 0.944 respectively; Figure 6). The Greek and the Bulgarian *Q. petraea* populations were classified in the same cluster (proportion of memberships 0.968 and 0.946 respectively). Similarly, in *Q. robur* the Greek and the Bulgarian populations were sorted to the same cluster (proportion of memberships 0.967 and 0.918 respectively). Interestingly, the statistic ΔK presented a secondary peak for K= 4, supporting an additional structure among provenances (Figure 4). By setting K≥5 we were not able to receive interpretable results.

Second, we analyzed population structure using the 'species discriminant' loci QrZAG30, 96 and 112, which showed high interspecific differentiation as demonstrated by $F_{ST}$ values. By setting K= 2 we were able to receive high membership proportions of the two species with the three *Q. petraea* populations varying between 0.954 and 0.984 and *Q. robur* varying between 0.954 and 0.983 (Figure 5). At the individual level, all but nine individuals (3.1%) could be assigned to one group with membership proportions higher than 0.85. These individuals were the same during all ten runs performed (results not shown). Moreover, limited numbers of individuals possessed membership proportions between 0.85 and 0.95. The number of such individuals varied between 16 and 17 in *Q. petraea* and between four and six in *Q. robur*, with the results differing slightly among runs. For K= 3 and K= 4 structures were weaker, but generally followed the same patterns as when all markers were used (results not shown). Posterior probability as revealed by ΔK was again highest for K= 2 (Figure 4).

Third, we conducted analysis using loci QpZAG1/5, 15, 110, QrZAG87 and 101 which exhibited higher differentiation within species than between the two species. Results appeared to be remarkably different than in the former two cases. For K= 2, the detected genetic structures did not correspond to the two species, but more to the different geographic regions. Both *Q. petraea* and *Q. robur* from Germany displayed relatively high proportions of membership to the same subpopulation (0.942 and 0.940; Figure 5). Respectively, *Q. petraea* and *Q. robur* from Greece could be classified to the same inferred "Balkan" cluster with values between 0.928 and 0.908. The population of *Q. petraea* from Bulgaria showed also high affinity to the same cluster (0.938), while Bulgarian *Q. robur* was more admixed and showed a membership proportion of 0.717 to the same cluster. A higher degree of admixture was also detected from the analysis of individual probabilities. On average, thirty nine individuals possessed admixed genotypes with membership proportions below 0.85 (results not shown). Results for K≥3 were not informative (Figure 6). Posterior data probability was highest for K=2, suggesting that these five loci share common geographic structures between the two species, while interspecific differentiation is not detectable (Figure 4).

4. DISCUSSION

Interspecific and intraspecific differentiation patterns in the closely related and interfertile *Q. petraea* and *Q. robur* were investigated using a highly informative set of microsatellite loci. We were able to discriminate clearly between the two species using populations from different regions – refugial and non-refugial. Furthermore, genetic structures within species were revealed from our data, providing evidence about provenance differentiation and local common structures between the two species, based on a group of 'provenance discriminant' loci. By partitioning our set of loci we could detect distinct diversity patterns for specific groups of loci. This might reflect the differential effects of speciation and local adaptation processes across the genome of both species.

Observation of the genetic structures with use of 14 microsatellite loci distinctly separated our samples to the species categories. Additional substructures were revealed within the two main species groups. However, splitting into main species structures and regional substructures was the result of the differential contribution of 'species' and 'provenance discriminant' loci. Partitioning of our marker set gave insights into the aforementioned differentiation components.

A common feature of the three 'species discriminant' loci is the high frequency of one allele in only one of the two species, resulting to high interspecific $F_{ST}$ values. Fixation of alleles favoured by selection and variability reduction at neighbouring hitchhiked genome areas is well recognized as the result of directional selection for a specific adaptive trait (Andolfatto 2001; Futuyma, 1998). Allele fixation can also be the result of genetic drift. However, diversity reduction due to genetic drift affects the genome rather uniformly (Schlötterer 2002). In our case, allele "135" at locus QrZAG96 shows over 80% frequency in all *Q. robur* populations compared to 2.1% over all *Q. petraea* populations, where the locus remains highly variable. On the other hand, loci QrZAG30 and 112 display high frequencies of a specific allele in *Q. petraea* remaining variable in *Q. robur*. In general, selection for one adaptive trait leads to hitchhiking, thus affecting the allele frequencies of neighbouring neutral marker loci in a wider genome area (Andolfatto, 2001). Interestingly, the $F_{ST}$-outlier locus QrZAG96 resides within a genomic area associated with a morphological QTL (petiole length as a ratio of the total leaf and petiole length), which is strongly discriminative between *Q. petraea* and *Q. robur* (Kremer et al., 2002; Saintagne et al., 2004). Similar hitchhiking effects might change diversity at loci QrZAG30 and 112 as well, causing reduction of heterozygosity and high frequencies of specific alleles. At locus QrZAG30 allele "170" was the most frequent in *Q. petraea* (56.7%), while it did not occur in any of the *Q. robur* individuals. We point out that frequency patterns of our 'species discriminant' loci are highly consistent throughout our study area, although ecological conditions and historical events are significantly different between the regions examined. The three loci apparently represent genome areas accounting for maintaining species integrity.

The aforementioned results agree with the findings of many previous studies supporting that differentiation between *Q. petraea* and *Q. robur* is restricted to some specific genome regions, while, in contrast, the rest of the genome is "permeable" and subjected to interspecific gene flow (Bodénès et al., 1997b; Gömöry et al., 2001; Scotti-Saintagne et al., 2004a). Restrictive use of the above three markers for carrying out Structure analysis in our study has been proved to be very effective in order to distinguish between species. Membership proportion to inferred groups corresponding to species using only the three 'species discriminant' microsatellite loci (QrZAG30, 96 and 112) was almost as high as the respective by using all of fourteen loci.

The same genomic pattern of differentiation occurs in many examples of interfertile species from the plant and the animal kingdoms. Selection against migrants at specific QTLs hampers the exchange of genetic information at these loci and hitchhiked areas around them, where $F_{ST}$ outlier loci can be found. Wang et al. (1997) and Nagy (1997) present results from hybrid zones among hybridizing *Artemisia* and *Gilia* subspecies, respectively, which are adapted to divergent environments and where hybrids demonstrate limited adaptability. Via and West (2008) described the same patterns in two partially reproductively isolated host races of pea aphids and revealed that hitchhiked areas can be extended up to more than 10cM. On the other hand, it has been shown that the rest of the genome of interfertile species can be susceptible to free exchange of neutral or mutually beneficial allelic variants through gene flow between them leading to common adaptations (Barton and Gale, 1993).

In our study, use of loci that displayed high intraspecific differentiation for Bayesian analysis demonstrated common structures between neighbouring populations of the two species. Clusters derived from five 'provenance discriminant' loci corresponded to the German and the Balkan provenances and no matching of the clusters to the two species could be observed. Differential adaptation to geography and ecological gradients might have contributed to the formation of common genetic structures between *Q. petraea* and *Q. robur*. Variation clines in oak have been described among others for bud burst, growth and stem form, and frost hardiness (Kremer et al., 2002). Similar variation patterns in adaptive genes have also been described in forest trees. For instance, Ingvarsson et al. (2006) in their study of *Populus tremula* found clinal variation within the putative gene Phy2, which is implicated in short-day induced bud set and growth cessation. Loci QpZAG1/5, 15, 110, QrZAG87 and 101 in our study possibly reflect local adaptive responses to environmental clines. QTL association studies reveal that locus QrZAG87 resides in a genome region coding for timing of bud burst (Scotti-Saintagne et al., 2004b). Moreover, Porth et al. (2005) could map a putative osmotic stress modulated gene (1T21) only 3,1cM apart from locus QrZAG101. At least two of the aforementioned adaptive traits are expected to follow diversity patterns corresponding to differential ecological conditions, with bud burst being influenced by photoperiod and osmotic stress being more intensive with increasing aridity. The fact that interspecific population pairs from ecologically equivalent areas in our study show minimal differentiation makes the hypothesis of common adaptation in presence of interspecific gene flow plausible.

By and large, our results support a scenario of directional selection at a limited number of genes and adjacent genome regions, which maintains species integrity. This is demonstrated by genetic structures based on interspecific highly differentiated loci ('species discriminant loci'). High degree of individual admixture and loose structure for a series of loci with neutral behaviour (MSQ13, QpZAG9, 16 and 104; QrZAG7 and 11) suggests that gene flow plays a homogenizing role at genome regions which are not liable to natural selection. Given that the two species have widely exchanged cpDNA haplotype variants through recent interspecific introgression, it is entirely plausible that they are able to exchange also selectively neutral nuclear variants. Under the same scenario, dominance of mutually beneficial variants might have led to common latitudinal adaptations, as expressed by the results from the 'provenance discriminant' loci in our study. Inclusion of populations from ecologically equivalent areas of other refugia would shed more light to this aspect. A further perspective is to extend the study using more marker loci in order to understand better the genomic architecture within and between the two species.

## 5. CONCLUSIONS

In conclusion, by using 14 nuclear microsatellite loci we were able to provide diverse information about differentiation patterns within and between *Q. petraea* and *Q. robur using* populations from Central Europe and the Balkans along an ecological gradient with increasing aridity towards south. The inclusion of the refugial area of the Balkan Peninsula, which harbours significant levels of biodiversity, is considered as an additional asset of this research. Restrictive use of two groups of loci – 'species discriminant' and 'provenance discriminant' – allowed discrimination of both species and provenances. On one hand, the two species could be well separated using only three 'species discriminant' loci. On the other hand, analysis of the five 'provenance discriminant' loci gave a completely different pattern of genetic structures, which corresponded only to local differentiation and included no taxonomic signal. Our results are in concordance with neo-Darwinian views of speciation, which describe the gene or interacting genes being the unit of speciation, with directional selection leading to two opposite directions (Wu, 2001; Wu and Ting, 2004; Via and West, 2008). 'Species discriminant' SSRs in our study seem to represent adaptive genomic areas, which are particular for each *Q. petraea* and *Q. robur*. Fixation of one allele is an obvious result of directional selection. 'Provenance discriminant' loci showed genomic sharing and common geographic structures between the two species, with latitudinal variation, possibly as a result of coadaptation. More extensive studies would be necessary to provide results with higher resolution and higher probabilities. The remaining six SSR loci seem to express "permeable" genome regions subjected to interspecific gene flow, which is supported by the high degree of individual and population admixture.

Population studies that utilize adaptive marker loci (e.g. SNPs; Scotti-Saintagne et al., 2004a) have had a limited impact to the description of selection and adaptation processes in natural populations of *Q. petraea* and *Q. robur*. Highly variable nuclear microsatellite loci remain a very useful tool for studying genetic diversity and structure, as well as for gathering information about adaptive processes. Our study points towards the use of sets of microsatellite loci as effective tools in provenance and species identification, which reduce both time and cost of the analysis. Use of such 'species' and 'provenance discriminant' markers with high resolution and a probable adaptive background makes a significant contribution to management and conservation of the two species. Finally, focusing on microsatellite loci with differential frequency patterns could lead to a targeted study of adaptive loci, aiming to elucidate the underlying evolutionary mechanisms in natural populations.

ACKNOWLEDGMENTS

We owe special thanks to Radko Babakov, Nour Alhammoud and Barbara Madariaga for their contribution to the samplings and laboratory analyses. We gratefully acknowledge the assistance of Dimitris Tampakiotis, Petko Ignatov, Iordan Christosov, Hermann Schott and Klaus Winkler during collections. We thank Dr. Aki M. Höltken for stimulating discussions. This research was conducted in partial fulfilment for the doctorate of the Albert-Ludwigs University of Freiburg regarding the first author. Charalambos Neophytou was supported by PhD scholarships from DAAD and the State Baden-Württemberg.

FIGURE LEGENDS

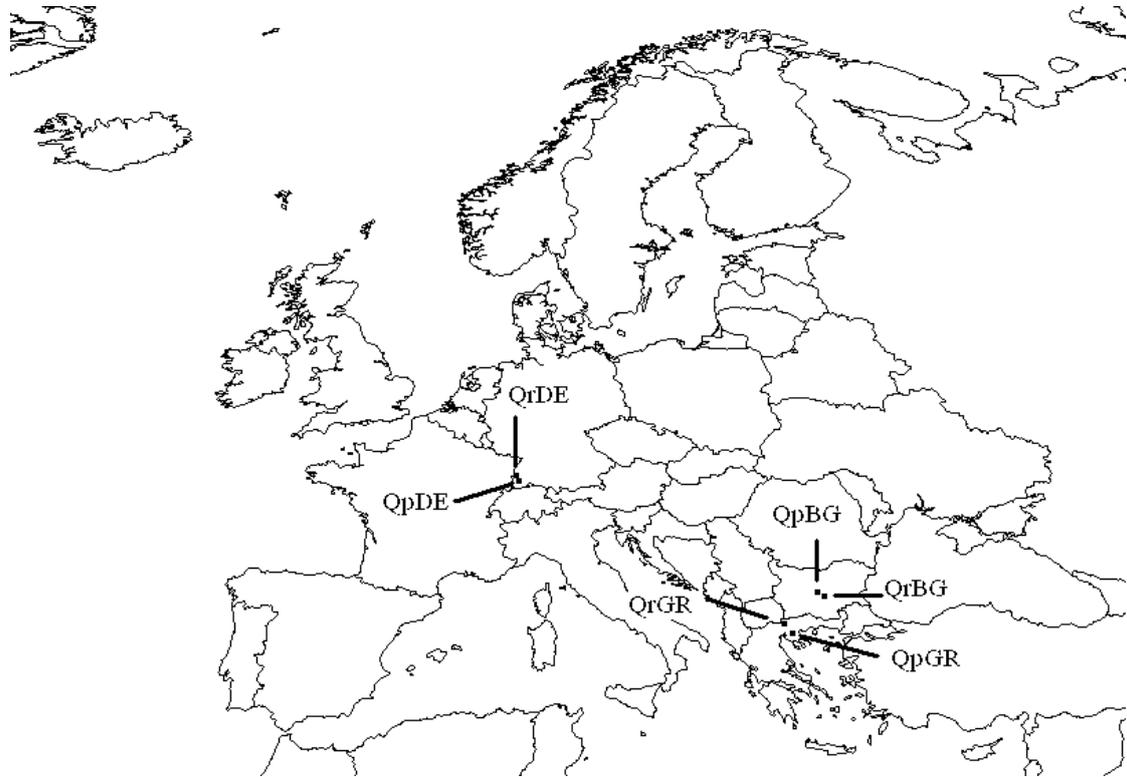

Fig. 1 – Locations of the sampled populations. Qp= *Quercus petraea*, Qr= *Quercus robur*; DE= Germany, BG= Bulgaria, GR= Greece.

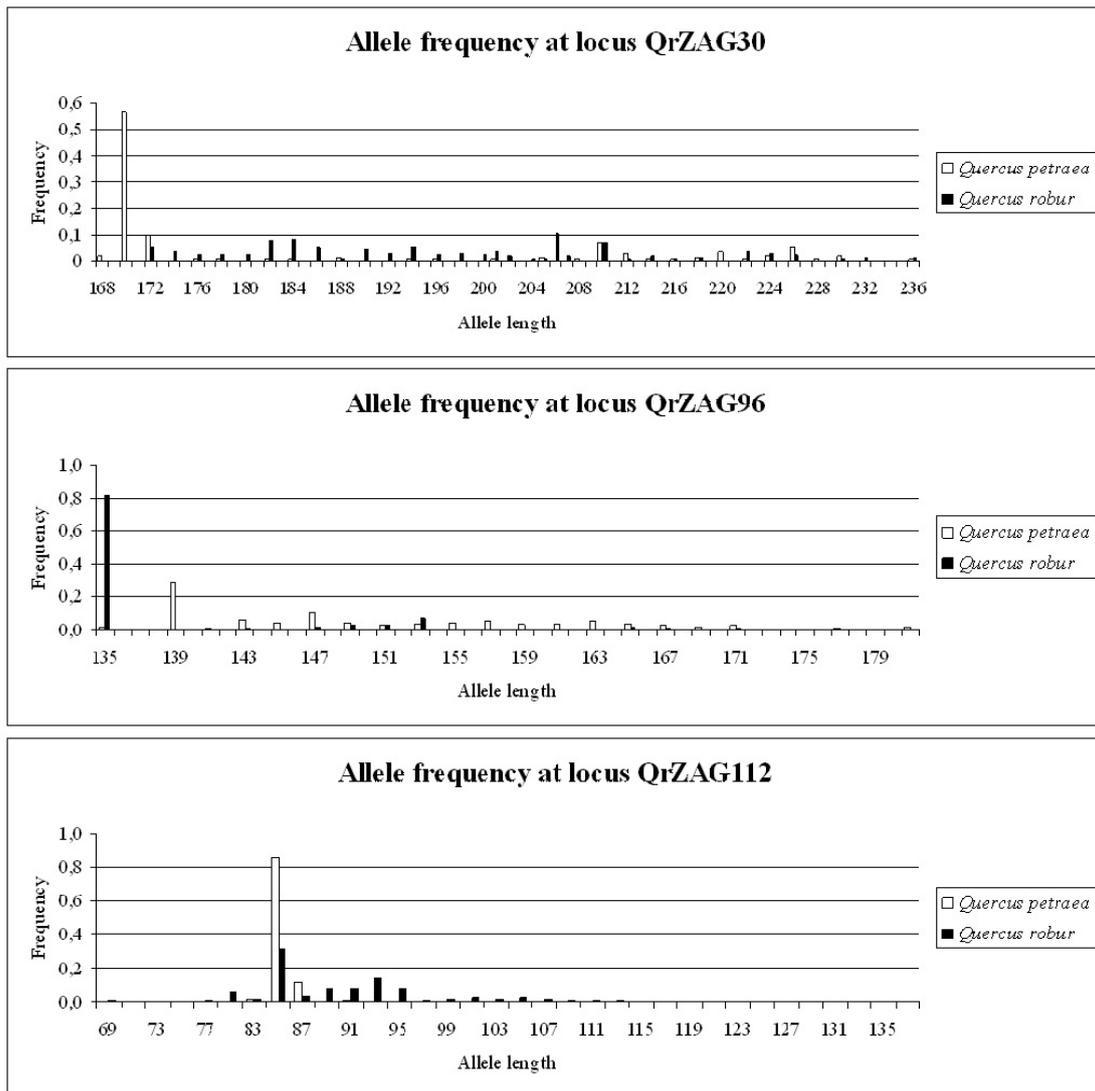

Fig. 2 – Allele frequency at loci QrZAG30, 96 and 112.
These loci exhibit high interspecific differentiation. For markers QrZAG30 and 112 alleles "170" and "85" occur with high frequencies in *Q. petraea* (56.7 % and 85.9 % respectively), while *Q. robur* shows a more diverse pattern. Allele QrZAG30 is completely absent from *Q. robur*. Conversely, at QrZAG96 allele "135" tends to be fixed in *Q. robur* (frequency 81.3%) while it occurs with a much lower frequency in *Q. petraea* (2.1%).

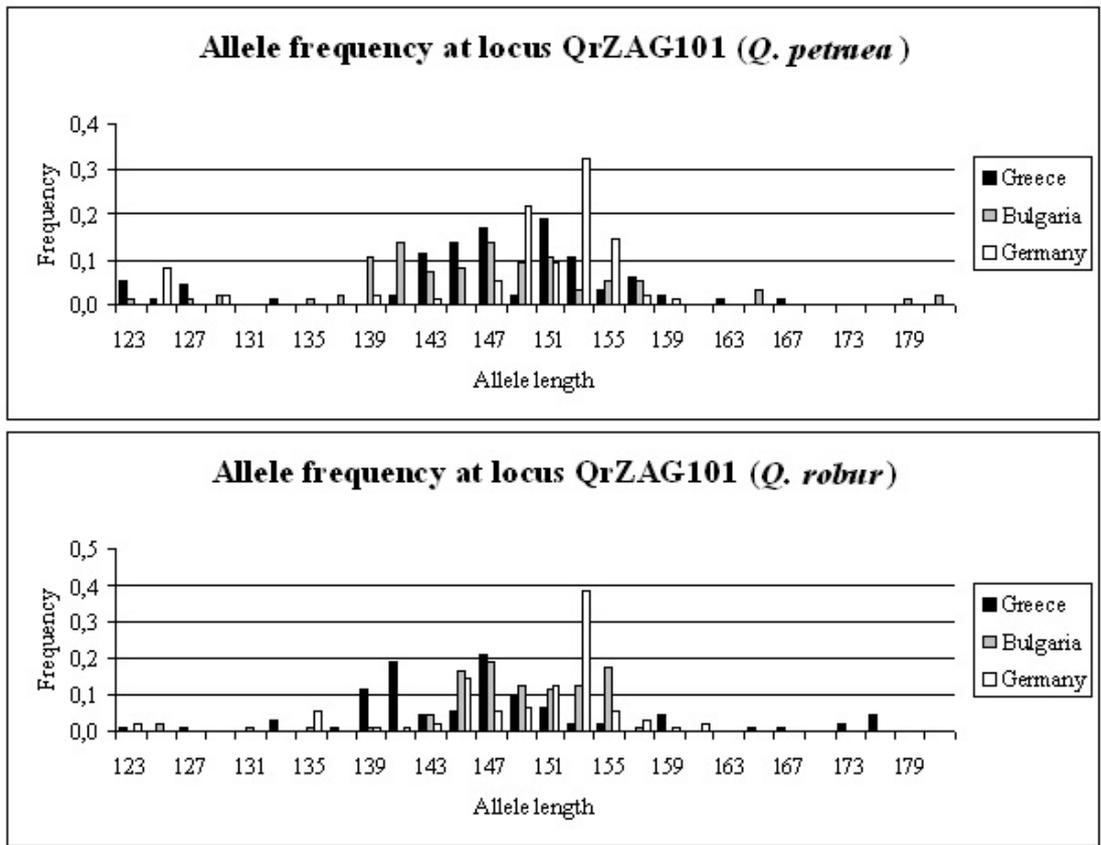

Fig. 3 – Allele frequency at locus QrZAG101.
Allele frequencies correspond more to the provenance than to the species. For example, allele "153" occurs with high frequency in both species in Germany, while this frequency decreases in Bulgaria and Greece.

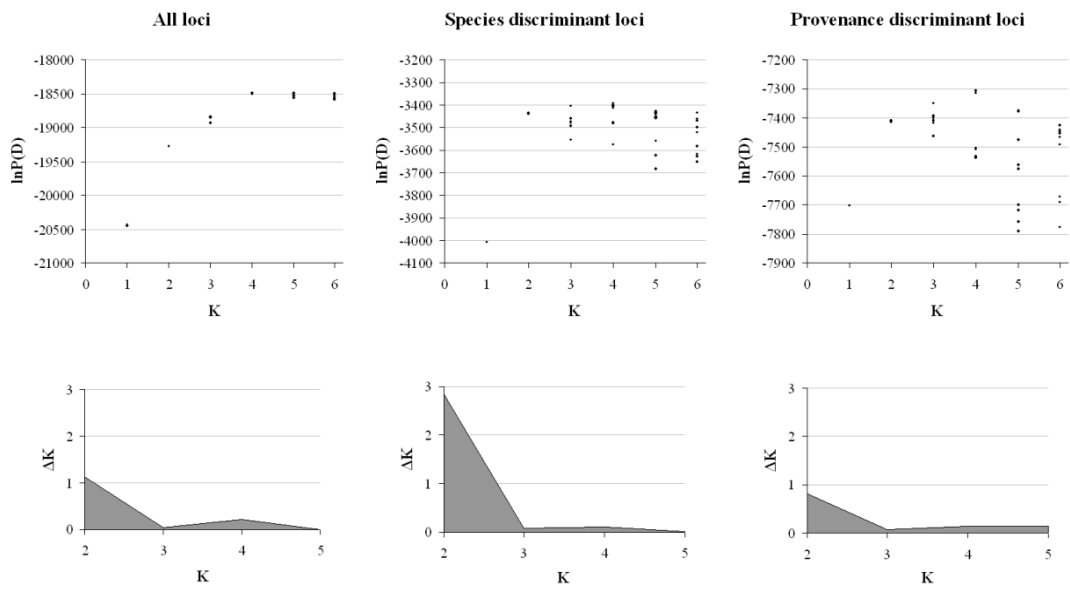

Fig. 4 – Data posterior probability among different Ks and values of the statistic ΔK. Posterior probability of data (ln(P|D)) for all runs (10 for each K) is presented. The highest ΔK corresponds to the uppermost level of hierarchy among runs with K= 1…6 (Evanno et al. 2005).

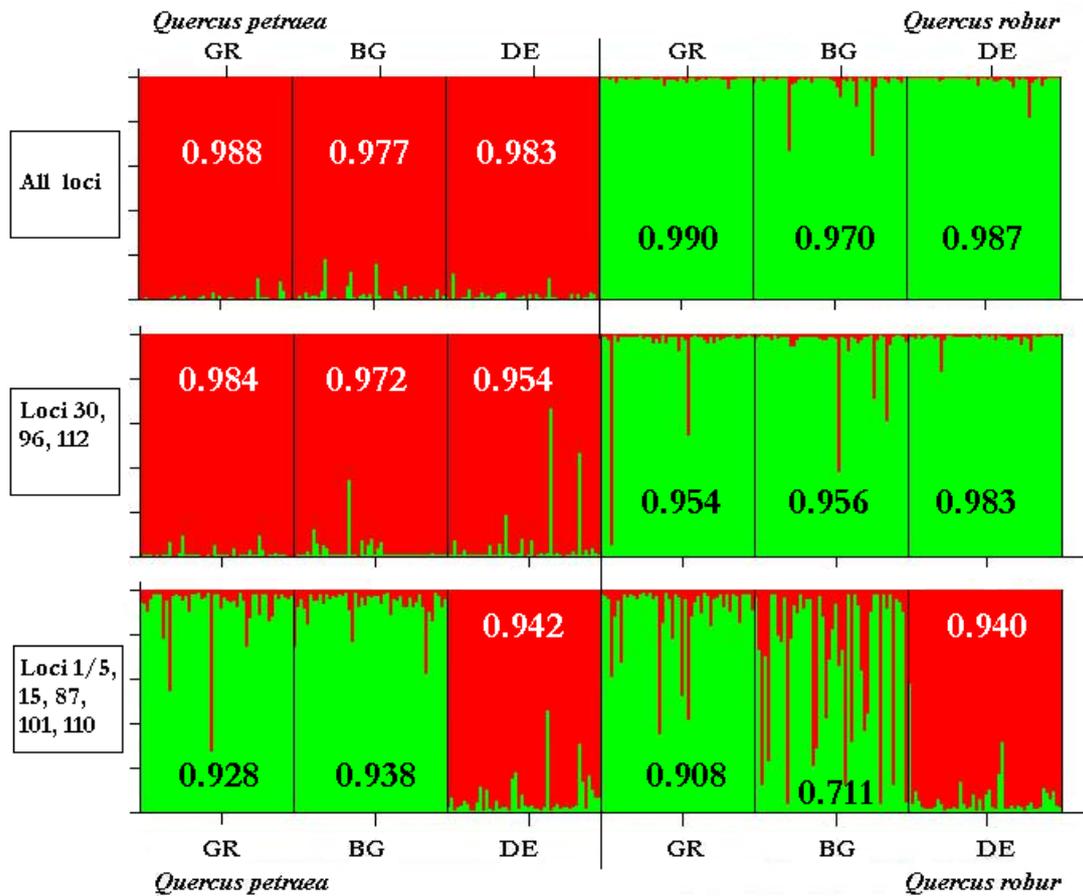

Fig. 5 – Genetic structures based different sets of loci.
Proportion of membership of each original population to two assumed subpopulations (K= 2) is presented. Graphs were based on the results from (a) All 14 analyzed SSRs (above), (b) Three 'species discriminant' SSRs with high interspecific $F_{ST}$s (QrZAG30, 96 and 112; middle) and (c) Five 'provenance discriminant' SSRs with high intraspecific $F_{ST}$s (QpZAG1/5, 16 and 110; QrZAG87 and 101; below)

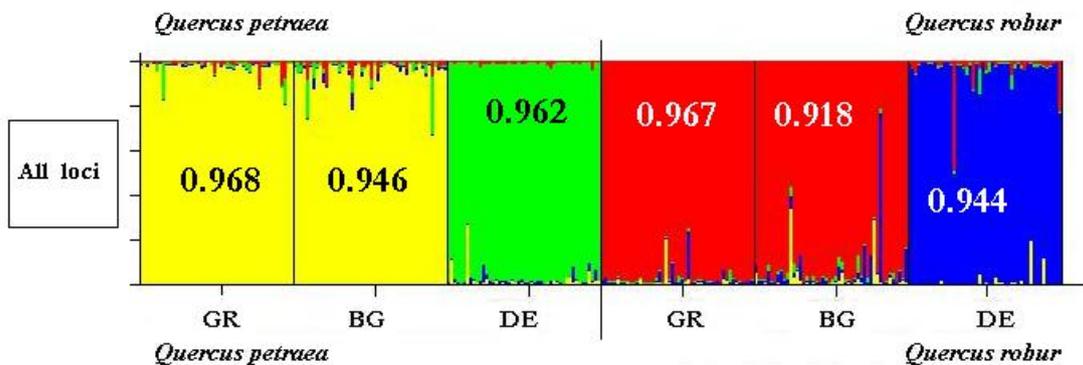

Fig. 6 – Genetic structures based on 14 SSRs for K= 2.
Proportion of membership of each original population to four assumed subpopulations (K= 4) based on the results from all 14 analyzed SSRs is presented.